\documentclass[pra,aps,twocolumn,showpacs]{revtex4}
\usepackage{times,epsfig,amssymb,amsfonts,amsmath}

\newcommand{\ket}[1]{|#1\rangle}
\newcommand{\bra}[1]{\langle #1|}
\newcommand{\proj}[1]{\ket{#1}\bra{#1}}

\begin{document}

\title{Multipartite quantum correlations in the frustrated and nonfrustrated regimes of a tunable triangular Ising system}
	
\author{Jun Ren$^1$}
\author{Fang-Man Liu$^1$}
\author{Li-Hang Ren $^1$}
\author{Z. D. Wang$^2$}
\email{zwang@hku.hk}
\author{Yan-Kui Bai$^{1,2}$}
\email{ykbai@semi.ac.cn}

\affiliation{$^1$ College of Physics and Hebei Key Laboratory of Photophysics Research and Application, Hebei Normal University, Shijiazhuang, Hebei 050024, People's Republic of China\\
$^2$ Guangdong-Hong Kong Joint Laboratory of Quantum Matter, Department of Physics, and HKU-UCAS Joint Institute for Theoretical and Computational Physics at Hong Kong, The University of Hong Kong, Pokfulam Road, Hong Kong, People's Republic of China}

\begin{abstract}
We study the multipartite quantum correlation (MQC) in a quantum transverse Ising system with the tunable triangular configuration, where it is found that the MQC itself cannot always discriminate the frustrated and nonfrustrated regimes of the ground state but the MQC combined with our newly defined MQC susceptibility can complete the task. Meanwhile, we reveal that the spatially anisotropic coupling is an effective and feasible tool for the MQC modulation in the ground state of frustrated Ising spins. Furthermore, we analyze the multipartite correlation properties in the thermal state, where it is shown that, unlike the thermally fragile MQC in the nonfrustrated regimes, there is a three-way trade-off relation among high MQC, strong thermal robustness, and the spatially anisotropic interactions in the frustrated spins. In addition, an experimental scheme for the MQC modulation via the anisotropic coupling is discussed in the system of cold atoms trapped in an optical lattice.

\end{abstract}


\maketitle

\section{Introduction}
Quantum correlation in the ground state of many-body systems is a kind of important physical resource for quantum information processing \cite{amico08rmp,horod09rmp,modi12rmp}. In particular, frustrated spin systems have received a lot of attention since it has highly degenerate ground states and can lead to multipartite quantum correlations (MQCs) \cite{sach99book,chiara18rpp}. The simplest spin frustration can be realized by three spins in a triangular lattice with antiferromagnetic interactions \cite{binder86rmp}, where the spins cannot order in the favored antiparallel pattern and the competition between interactions and lattice geometry results in the degenerate manifold of ground states. Recently, theoretical and experimental studies have shown that the spatially anisotropic frustrated triangular structures have exotic phases and potential applications in quantum computation \cite{Yoshida15np,Zhang16prl,Ito16prb,Keles18prl,Tala20prapp}.

In many-body quantum systems, the ground state can be in different phases at zero temperature \cite{sach99book}, and the MQC is an effective tool to characterize the properties of quantum phases \cite{weitc05pra,olivei06pra,olivei06prl,facchi10njp,mon10pra,giampa13pra,sunzy14pra,hofm14prb,Bayat17prl,Yamasaki18pra,haldar20prb}. Entanglement monogamy in multipartite systems means that quantum entanglement cannot be freely shared among different subsystems \cite{bennett96pra} and the quantitative monogamous relations can be used to constitute the measures or indicators for genuine MQC \cite{coffman00pra,osborne06prl,christandl04jmp,fan07pra,byw07pra,kimjs09pra,byw09pra,kimjs10jpa,cornelio10pra,bxw14prl,songbai16pra,ren21npj}. Utilizing the tripartite quantum correlation based on the monogamy relation of negativity \cite{fan07pra,vidal02pra}, Rama Koteswara Rao \emph{et al} studied the ground state of quantum transverse Ising model in a fixed triangular configuration via the nuclear magnetic resonance (NMR) system \cite{rkrao13pra}. They found that, unlike the bipartite quantum correlation, the multipartite quantum correlation can be used to distinguish between the frustrated and the nonfrustrated regimes in the ground state of the three spins system, where the MQC in the nonfrustrated regime exhibits the higher values than that in the frustrated one. It is noted that the ground state property depends greatly on the anisotropy of the system, such as anisotropic Heisenberg chains \cite{wang01pla,kamta02prl,zhou03pra,zhang05pra} and spatially anisotropic triangular lattices \cite{yunoki06prb,eckardt10epl,hauke13prb}. In comparison with anisotropic Heisenberg models, the spatially anisotropic lattices are more feasible and, in recent years, experiments on different physical platforms (such as trapped ions, cold atoms, and superconducting systems) have realized this kind of couplings among spins in the triangular configuration  \cite{kim10nat,Struck11sci,wang18natnano,niskanen07sci,Ryazanov01prb,Grosz11prb,Kosior13pra,Seo13prl}. Therefore, it is desirable to study the ground state properties in a tunable spatially anisotropic system and investigate a more general relation between the MQC and the frustration of ground state in Ising spins. Meanwhile, it is also worthy of further analyzing the MQC at finite temperatures, since it plays a crucial role in quantum computation \cite{cory98prl,knill98prl,rauss01prl,datta08prl,lanyon08prl,monz09prl,lanyon09natp}.

\begin{figure*}
	 \epsfig{figure=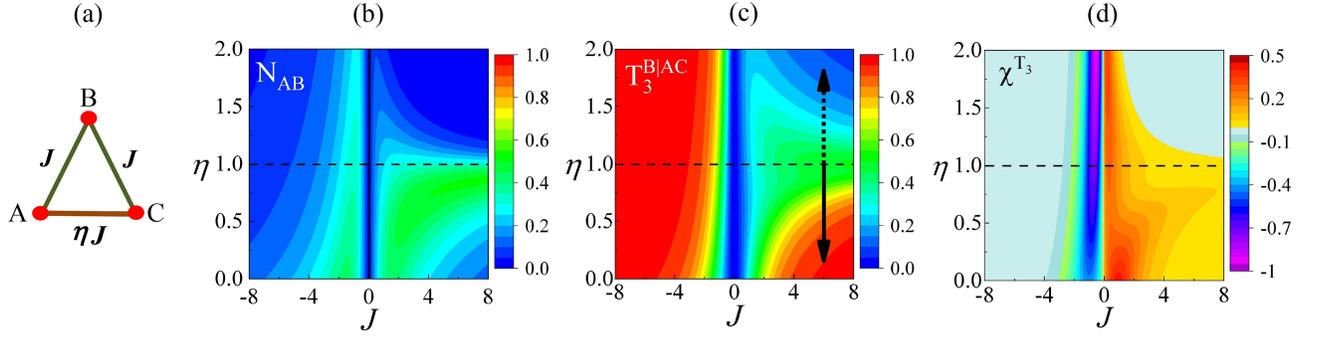,width=0.95\textwidth}
\caption{(Color online) (a) The schematic diagram of the spatially anisotropic triangular Ising model. (b) The negativity $N_{AB}$ of ground state as a function of the coupling strength $J$ and the tunable anisotropic parameter $\eta$. (c) The change of tripartite quantum correlation $T_{3}^{B|AC}$ in the ground state along with the parameters $J$ and $\eta$, where the solid and dotted arrows indicate two tunable directions of $\eta$ from isotropy to anisotropy. (d) The MQC susceptibility $\chi^{T_3}=\partial T_{3}^{B|AC}/\partial J$ as a function of the parameters $J$ and $\eta$.}
\end{figure*}

In this paper, we study the properties of MQC in a spatially anisotropic triangular Ising model with tunable couplings, where the aim of our work is triplex. First, we focus on genuine MQC and establish a general link between the MQC and the ground state of frustrated Ising spins in a tunable triangular configuration, which generalizes the case for the fixed triangular configuration in the NMR system \cite{rkrao13pra}. Second, we analyze the effect of spatial anisotropy on the MQC of the ground state, and explore an effective and feasible method for the MQC modulation under current experiment technologies. Third, we investigate the MQC in the thermal state, and explore the inherent relation among three key contents, i.e., high multipartite correlations, strong thermal robustness and spatial anisotropy in the frustrated spins. The paper is arranged as follows. In Sec. II, we study the MQC in the ground state of frustrated Ising spins, where it is revealed that the MQC combined with our newly defined MQC susceptibility can distinguish the frustrated and the nonfrustrated regimes. Meanwhile, we find the spatially anisotropic coupling is an effective tool for the MQC modulation in the frustrated regime. In Sec. III, we investigate the MQC in the thermal state, where we find that there exists a three-way trade-off relation among high MQC, strong thermal robustness, and the spatially anisotropic interactions in the frustrated regime of the tunable Ising system. Finally, in Sec. IV, we discuss the realization of the MQC modulation in the system of cold atoms, analyze two typical spatially anisotropic configurations possessing relatively high MQC and well thermal robustness at the same time, and give a brief conclusion.

\section{multipartite quantum correlation in the ground state of frustrated Ising spins}

\subsection{The MQC in a triangular Ising model with one tunable asymmetric interaction}

We first consider the spatially anisotropic triangular Ising system with one tunable asymmetric interaction, for which the Hamiltonian has the form
\begin{equation}\label{1}
H=J(\sigma_A^x \sigma_B^x+\sigma_B^x\sigma_C^x+\eta \sigma_A^x \sigma_C^x)+h(\sigma_A^z+\sigma_B^z+\sigma_C^z),
\end{equation}
where $J$ is the coupling strength of interaction with the nonnegative $\eta$ being a tunable asymmetric parameter, $h$ is the strength of the transverse field, $\sigma_i^x$ and $\sigma_i^z$ are Pauli operators with $i=\{A, B, C\}$ being three spins located at the vertices of a triangular lattice as shown in Fig. 1(a). When $J/h>0$, the system possesses antiferromagnetic interactions and is frustrated, while, in the case of $J/h<0$, the interactions are ferromagnetic and the system is nonfrustrated. In this paper, we set the parameter $h=1$ for simplicity. The spatially anisotropic triangular configuration with two equally coupled legs is equivalent to a three-spin chain with the nearest-neighbor interaction $J$ and the next-nearest-neighbor interaction $\eta J$. Experimentally, this kind of tunable anisotropic triangular coupling has been realized in ion traps \cite{kim10nat} and optical lattices \cite{Struck11sci}.

In order to study the MQC in the tunable system, we solve the Hamiltonian in Eq. (1).
After some derivations, we can obtain the concise expression of the ground state up to local unitary transformations
\begin{equation}\label{2}
\ket{\psi_0}_{ABC}= (\alpha\ket{001}+\gamma\ket{010}+\alpha\ket{100}+\chi\ket{111})/K_0,
\end{equation}
where $K_0=[2\alpha^2+\gamma^2+\chi^2]^{1/2}$ is the normalization coefficient with the amplitudes $\alpha$, $\gamma$, and $\chi$ being the functions of the coupling strength $J$ and the tunable asymmetric parameter $\eta$, for which the analytical expressions are presented in Appendix A. In fact, the ground state in the frustrated regime is the superposition of six degenerate states and the one for the nonfrustrated case is the superposition of two states with the ferromagnetic order. In Eq. (2), the ground state belongs to the class of tetrahedral states \cite{Carteret00jpa,Brun01pla}, which is a kind of multipartite quantum resource states and can be used to perform the controlled teleportation \cite{Li14pra}.

Entanglement negativity \cite{vidal02pra,plenio05prl,zyc98pra} is an important quantum correlation measure, which is computable for bipartite mixed states and is defined as \cite{vidal02pra}
\begin{equation}\label{3}
N(\rho_{AB})=\|\rho_{AB}^{T_A}\|_1-1,
\end{equation}
where $\|\cdot\|_1$ denotes the trace norm, which is the sum of the moduli of eigenvalues for the partial transposed matrix $\rho_{AB}^{T_A}$. For three-qubit systems, the genuine multipartite quantum correlation can be characterized by the residual correlation according to the distribution of squared negativities. Ou and Fan proved the monogamy property in pure states and defined the measure $\pi_3^{A|BC}= N_{A|BC}^2-N_{AB}^2-N_{AC}^2$ \cite{fan07pra}. Here, in order to make the value of tripartite quantum correlation comparable with bipartite negativities, we make use of a modified tripartite quantum correlation measure
\begin{equation}\label{4}
T_{3}^{A|BC}(\ket{\Psi})=[N_{A|BC}^{2}-N_{AB}^{2}-N_{AC}^{2}]^{\frac{1}{2}},
\end{equation}
where $N_{A|BC}$ is the bipartite negativity between subsystems $A$ and $BC$, $N_{Aj}$ the two-qubit negativity with $j=\{B,C\}$, and the superscript $A|BC$ of $T_3$ means that subsystem $A$ is the central qubit in the tripartite correlation.

Utilizing the above measures, we can analyze quantum correlations in the ground state of Eq. (2). For the two-qubit subsystem $\rho_{AB}=\mbox{tr}_{C}[\proj{\psi_0}]$, we obtain the negativity
\begin{equation}\label{5}
N_{AB}(J,\eta)=(m-\alpha^2-\gamma^2)/K_0^2,
\end{equation}
where $m=[(\alpha^2-\gamma^2)^2+4\alpha^2\chi^2]^{1/2}$. In Fig. 1(b), the negativity $N_{AB}$ is plotted as the function of the interaction parameters $J$ and $\eta$, where the two-qubit negativities in both nonfrustration and the frustration regimes have small values whose relative differences between the two regimes are not distinct, and thus $N_{AB}$ may not be suitable for distinguishing the two regimes of the ground state, which is similar to the case of fixed configuration in Ref. \cite{rkrao13pra}. The dashed line in the figure indicates the isotropic coupling with $\eta=1$.

For the MQC in the ground state $\ket{\psi_0}_{ABC}$, we use the tripartite correlation $T_3^{B|AC}$ and take the qubit $B$ as the central one according to the triangular configuration of the tunable Ising model. After some derivations, we can get
\begin{equation}\label{6}
T_{3}^{B|AC}(\ket{\psi_0})=2 \sqrt{2}[\left(\alpha^{2}+\gamma^{2}\right) m-\left(\alpha^{2}-\gamma^{2}\right)^{2}]/K_{0}^{2},
\end{equation}
where the parameters $K_0$ and $m$ have the same expressions as those in Eqs. (2) and (5), and the multipartite correlation is also the function of parameters $J$ and $\eta$. In Fig. 1(c), we plot the change of $T_{3}^{B|AC}$ along with the coupling strength $J$ and the anisotropic parameter $\eta$. In the red region away from the critical point ($J=0$), $T_{3}^{B|AC}$ in the nonfrustration regime has a relatively large value and the anisotropic parameter $\eta$ has little influence on the correlation, while the MQC in the frustration regime ($J>0$) is sensitive to the anisotropic parameter $\eta$. As indicated by the two arrows in Fig. 1(c), $T_{3}^{B|AC}$ increases along with the direction of the solid arrow when $\eta\leq 1$, but the MQC decreases along with the direction of the dotted arrow when $\eta\geq 1$. In comparison to the case for fixed configuration in NMR systems \cite{rkrao13pra}, it is obvious that the MQC in the tunable Ising system cannot always distinguish the nonfrustrated and the frustrated regimes. We find that, when the anisotropic parameter $\eta\geq 1$, and the MQC away from the critical point can be served as an effective indicator to distinguish the two regimes. However, when $\eta< 1$ or $J$ close to critical point, the tripartite quantum correlation itself may not be suitable for the discrimination, because $T_{3}^{B|AC}$ in both regimes are comparable as shown in Fig. 1(c).

In order to give a more general link between the MQC and the ground state of the tunable Ising system, we can further define a new multipartite correlation indicator, \emph{i.e.}, the MQC susceptibility which has the form
\begin{equation}\label{7}
\chi^{T_3} \equiv \partial T_{3}^{B|AC}/\partial J,
\end{equation}
which is the first-order derivative of the tripartite correlation versus coupling strength $J$. The definition of $\chi^{T_3}$ is similar to that of coherence susceptibility given in Ref.~\cite{chen16pra}. As shown in Fig. 1(d), when the asymmetric parameter $\eta <1$ (the tripartite quantum correlation $T_{3}^{B|AC}$ loses its efficacy), the susceptibility $\chi^{T_3}$ can distinguish the nonfrustration regime since $\chi^{T_3}$ has the larger value in the frustration one. Moreover, when the coupling strength $J$ is close to 0, the $\chi^{T_3}$ is negative in the case $J<0$ but positive in the case $J>0$. Therefore, in the tunable system with the Hamiltonian in Eq. (1), the MQC itself cannot always characterize the ground state, which is quite different from the case of fixed triangular configuration where the MQC in the frustrated regime always has the lower value than that of the nonfrustrated one\cite{rkrao13pra}. With the help of the newly defined $\chi^{T_3}$, we establish a more general link between the MQC and the ground state in the tunable system where the $T_{3}^{B|AC}$ combined with its susceptibility $\chi^{T_3}$ constitute a set of effective tools to discriminate the frustration ($J>0$) and the nonfrustration ($J<0$).

\begin{figure}
	 \epsfig{figure=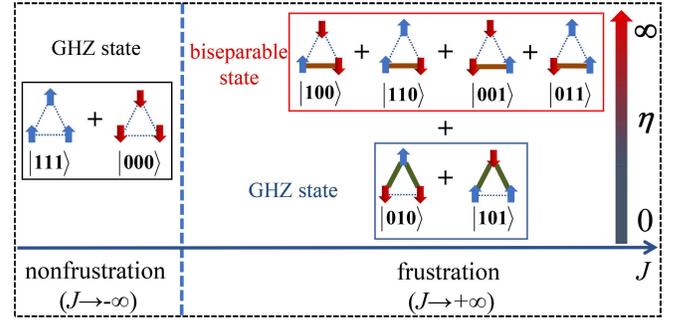,width=0.48\textwidth}
	\caption{(Color online) The schematic diagram for the relationship between the spatially anisotropic coupling and quantum correlation structure in the ground state far away from the critical point.}
\end{figure}

The influence of the spatially anisotropic parameter $\eta$ on tripartite quantum correlation $T_{3}^{B|AC}$ is different in the frustrated and the nonfrustrated regimes as shown in Fig. 1(c). Here, we analyze the reasons and consider the regions far away from the critical point. In the nonfrustrated regime with ferromagnetic interaction, regardless of the anisotropic strength, the ground state of the system always tends to the state of three spins up or down at the same time. In the case of $J\rightarrow -\infty$, the ground state is just the GHZ state with the maximally tripartite quantum correlation as shown in the left part of Fig. 2. Therefore, the anisotropic parameter $\eta$ does not change the property of ground state and has little influence on the MQC in the nonfrustration regime with the larger value of $J$. However, the case for the frustrated regime is quite different as shown in the right part of Fig. 2. When $\eta=1$ (namely isotropic structure) with the larger $J$, the ground state tends to the equal probability superposition of the six degenerate states \{$\ket{100}$, $\ket{010}$, $\ket{001}$, $\ket{011}$, $\ket{101}$, $\ket{110}$\}, for which, no matter how strong the coupling strength is, the tripartite quantum correlation $T_{3}^{B|AC}$ does not exceed $\sqrt{1/3}$ (the equality corresponds to $J \rightarrow \infty$). As the value of $\eta$ moves away from 1, the anisotropy will affect the competition among these degenerate states, thereby affecting the correlation structure of the ground state. These degenerate states are divided into two sets according to the value of $\eta$. When $\eta<1$, the pair interactions of A-B and B-C are stronger than that of A-C, which implies that the ground state tends to ensure the pair spins AB and BC being opposite, namely $\ket{101}$ and $\ket{010}$. The superposition of the two degenerate states with equal probabilities ($\eta \rightarrow 0$) is just a three-qubit GHZ state with maximal tripartite correlation. Conversely, in the case of $\eta>1$, the interaction between A and C dominates and the ground state tends to be the equal probability superposition of the four degenerate states $\ket{100}$, $\ket{110}$, $\ket{001}$ and $\ket{011}$ (with opposite spins between A and C), which is a biseparable state when $\eta\rightarrow \infty$ and its tripartite quantum correlation $T_{3}$ is zero. The above analysis on the MQC indicates that, for the case of frustrated regime, the MQC is sensitive to the spatially anisotropic coupling (parameter $\eta$), which provides an effective strategy for multipartite correlation modulation and is feasible under current experiment technologies.

It is well known that the magnetic susceptibility $\chi^{\textbf{M}}$ is an important indicator of quantum phase transition in spin systems \cite{lou04prb,hong08prb,Sidorov11prb,Hotta18prb}, where $\chi^{\textbf{M}}=\partial \bra{\psi_0} \boldsymbol{M} \ket{\psi_0}/\partial J$ and  $\boldsymbol{M}=-g\mu_B \boldsymbol{\sigma}/2$ with $\boldsymbol{\sigma}=\sum_i \boldsymbol{\sigma}_i$ being total spin operator. Here, we compare the behaviors of the magnetic susceptibility and the MQC susceptibility $\chi^{T_3}$ in the tunable Ising system. For simplicity, we calculate $\partial \bra{\psi_0} \boldsymbol{\sigma} \ket{\psi_0}/\partial J$ which is equivalent to the $\chi^{\textbf{M}}$ up to a constant coefficient. In Fig. 3, the two susceptibilities $\chi^{T_3}$ and $\chi^{\textbf{M}}$ are plotted as functions of the coupling strength $J$ for some anisotropic configurations, where two susceptibilities have the similar behaviors along with the changes of $J$ and $\eta$. Both $\chi^{T_3}$ and $\chi^{\textbf{M}}$ are negative in the nonfrustrated regime and positive in the frustrated regime. Moreover, when $|J|$ is small, the magnetic susceptibility can also distinguish the two regimes in the tunable system. In experiment, the total magnetic moment is very small for few spin systems and thus may be difficult to measure, while the MQC susceptibility $\chi^{T_3}$ can be calculated by the reconstructed density matrix via quantum state tomography \cite{rkrao13pra,paris04lnp}.

\begin{figure}
	 \epsfig{figure=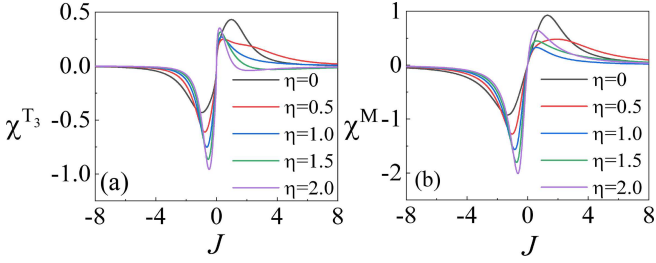,width=0.48\textwidth}
	\caption{(Color online) The susceptibilities $\chi^{T_3}$ and $\chi^{\textbf{M}}$ as functions of the coupling strength $J$ for some anisotropic triangular configuration with different values of $\eta$.}
\end{figure}

\subsection{The MQC in the ground state of completely spatially anisotropic triangular Ising systems}

We have studied the relation between the MQC and the ground state property in the spatially anisotropic triangular Ising system with one asymmetric interaction. In practical quantum systems, there also exist completely spatially anisotropic coupling interactions, such as the dipole-dipole interaction Heisenberg model \cite{Keles18prl}, the tunable artificial spin ice structure \cite{Tala20prapp}, and the superconducting qubit architecture \cite{Grosz11prb}, etc. Next, we analyze the completely spatially anisotropic triangular Ising system with three different couplings, for which the Hamiltonian reads
\begin{equation}\label{8}
H'=J(\sigma_A^x \sigma_B^x+\omega\sigma_B^x\sigma_C^x+\eta \sigma_C^x \sigma_A^x)+h(\sigma_A^z+\sigma_B^z+\sigma_C^z),
\end{equation}
where $J$ is the coupling strength with $\omega$ and $\eta$ being two tunable anisotropic parameters. The ground state for the above system  has also the form of tetrahedral state
\begin{equation}\label{9}
\ket{\psi'_0}=(\xi\ket{001}+\zeta\ket{010}+\delta\ket{100}+\tau\ket{111})/\mathcal{K}_0,
\end{equation}
where $\mathcal{K}_0$ is the normalization coefficient and the four amplitudes are different in general. As two examples, we choose the value of spatially anisotropic parameter $\omega$ to be 0.8 and 1.2, and calculate the tripartite quantum correlation $T_{3}^{B|AC}$ according to the ground state (the expressions for the coefficients and the MQC are presented in Appendix B).

\begin{figure}
	 \epsfig{figure=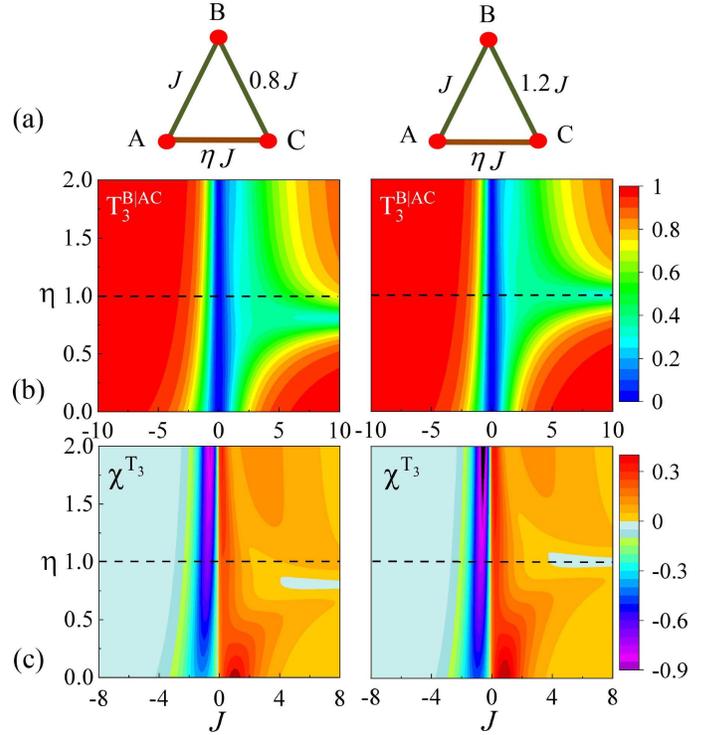,width=0.5\textwidth}
	\caption{(Color online) The MQC in the ground state of completely spatially anisotropic Ising system with three different interactions: (a) the schematic diagram for the interaction configurations $1:0.8:\eta$ and $1:1.2:\eta$, (b) the change of $T_{3}^{B|AC}$ along with anisotropic parameters $J$ and $\eta$, (c) the MQC susceptibility $\chi^{T_3}=\partial T_{3}^{B|AC}/\partial J$ as a function of parameters $J$ and $\eta$, where the left panels are the case of $\omega=0.8$ and the right ones for $\omega=1.2$.}
\end{figure}

In Fig.~4, we plot the MQC and the corresponding $\chi^{T_3}$ for the interaction configurations $1:0.8:\eta$ (left column) and $1:1.2:\eta$ (right column). In the nonfrustrated regime, the anisotropic parameters $\omega$ and $\eta$ have little influence on the tripartite quantum correlation as shown in Fig.~4(b), which is similar to that of the single anisotropic coupling in Fig.~1(c).
In the frustrated regime, when $\eta <1$, both the ground states for the two configurations tend to be a GHZ state and have the larger tripartite quantum correlation. However, when $\eta >1$ in the frustrated regime, the situation for the two configurations is different from that of the single anisotropic interaction, and the MQC can attain to a relatively large value. The reason is that, when $\eta>1$ and $\omega\neq 1$, the energies of the four degenerate states \{$\ket{100}$, $\ket{110}$, $\ket{001}$, $\ket{011}$\} are not the same and the equal probability superposition of the four states is destroyed ($\ket{100}+\ket{011}$ dominates for $\omega=0.8$, the other pair dominates for $\omega=1.2$), which results in a larger tripartite quantum correlation. Moreover, in the frustrated regime with the larger $J$, the anisotropic parameters $\omega$ and $\eta$ jointly determine the lower tripartite quantum correlation regions where the minimum corresponds to the configuration of anisotropic interactions being equivalent to the case of single asymmetric interaction, namely $1:0.8:0.8$ and $1:1.2:1$.

In the completely anisotropic systems, it is obvious that the tripartite quantum correlation $T_{3}^{B|AC}$ alone cannot distinguish between the nonfrustrated and frustrated regimes, but the properties of ground state can still be discriminated by the set of effective tools we introduced, \emph{i.e.}, the MQC and its susceptibility $\chi^{T_3}$ as shown in Figs.~4(b) and 4(c).

\section{The MQC in the thermal state of the tunable anisotropic Ising system}

In the past two decades, considerable efforts have been made to study the thermal bipartite entanglement or quantum correlation in spin systems and analyze the effects of magnetic field and temperature, see, for example, Refs. \cite{wang01pla,kamta02prl,zhou03pra,zhang05pra,arnesen01prl,wang01pra,wang02pra,sun05njp,abliz06pra,ww06pra,hide09prl,park17prl,Shim08prb} and the review papers \cite{amico08rmp,bera18rpp}. Since multipartite quantum correlation is a kind of important physical resource for quantum information processing \cite{vedral04njp,Nakata09pra,Campbell13njp,sun19pra}, it is necessary to investigate the situation for the MQC at finite temperatures. In particular, we will focus on the inherent relation among the high MQC, strong thermal robustness and spatially anisotropic couplings in the tunable Ising system with the triangular configuration.

The thermal state under the equilibrium at a finite temperature $T$ can be written as $\rho(T)=\frac{1}{Z} e^{-\beta H}$ in which $Z=\mbox{tr}[\mbox{exp}(-\beta H)]$ is the partition function with $H$ being the system Hamiltonian and  $\beta=1/k_{B}T$ (for simplicity, the Boltzmann constant is chosen to be $k_B=1$). In the basis of eigenvectors, the thermal state can be further expressed as
\begin{equation}\label{10}
\rho(T)=\frac{1}{Z} \sum_{i}e^{-E_i/T}\ket{e_{i}}\bra{e_{i}},
\end{equation}
where $E_i$s are energy eigenvalues of the given Hamiltonian and the coefficient $e^{-E_i/T}$ determines the proportion of the eigenstate $\ket{e_i}$ in the mixed state.

We first consider the isotropic triangular Ising system for which the Hamiltonian is described by Eq. (1) with the tunable parameter $\eta=1$. The expressions for energy eigenvalues and eigenvectors can be found in Appendix A. For the thermal state $\rho(T)$, we calculate the tripartite quantum correlation $T_3^{B|AC}$ and plot it as a function of temperature $T$ in Fig. 5(a), where the dotted and solid lines correspond to the cases of nonfrustrated ($J<0$) and nonfrustrated ($J>0$) regimes with different values of coupling strength. It is obvious that the MQC decreases along with the temperature $T$ in both cases. However, the correlation $T_3^{B|AC}$ in frustrated states ($J=2,4,6$) has the stronger thermal robustness than that in the nonfrustrated states ($J=-2,-4,-6$), even if the initial tripartite quantum correlations in nonfrustrated states have the larger values. In addition, the larger the coupling parameter $|J|$ is, the more rapidly the MQC decays in the nonfrustrated states; while the larger the coupling $J$, the higher the MQC in the frustrated states.

\begin{figure}
	 \epsfig{figure=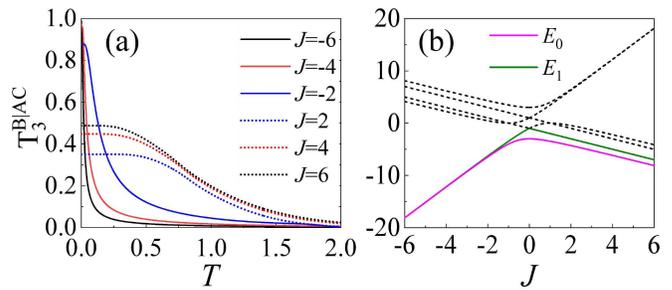,width=0.48\textwidth}
	\caption{(Color online) (a) The thermal effect on tripartite quantum correlation $T_{3}^{B|AC}$ in an isotropic triangular Ising model where the dotted lines correspond to the frustrated regime and the solid lines represent the cases of nonfrustrated regime. (b) The change of energy eigenvalues along with the coupling strength $J$ for the Hamiltonian of isotropic triangular Ising model, where two lowest eigenvalues $E_0$ and $E_1$ are highlighted by the solid lines.}
\end{figure}

In order to analyze the different thermal robustness of MQC in the nonfrustrated and frustrated regimes, we further plot the change of energy eigenvalue $E_i$ along with the coupling strength $J$ as shown in Fig. 5(b). According to the expression of thermal state in Eq. (10), we know that the mixed probability of component $\ket{e_i}$ is proportional to the value $-E_i$, and thus the eigenvector with the negative energy eigenvalue is dominant while the one with positive eigenvalue has a small weight in the thermal state for a given lower temperature $T$. As shown in the figure, when $J<0$ in the nonfrustrated phase, the energy eigenvalues $E_0=J-2\sqrt{J^2-J+1}-1$ (purple solid line) and $E_1=J-2\sqrt{J^2-J+1}+1$ (green solid line) are negative and other eigenvalues are positive (black dotted lines), which leads to the result that the thermal state is mainly comprised of the eigenvectors $\ket{e_0}$ and $\ket{e_1}$. The two eigenvectors have the forms
\begin{eqnarray}\label{11}
\ket{e_0}&=&(\ket{001}+\ket{010}+\ket{100}+a\ket{111})/K_0,\\
\ket{e_1}&=&(b\ket{000}+\ket{011}+\ket{101}+\ket{110})/K_1,
\end{eqnarray}
where $K_i$s with $i=0,1$ are normalization coefficients, and the parameters $a=(E_0+J-1)/(E_0+J+3)$, $b=(E_1-2J+1)/J$. Along with the increase of $T$, the nonfrustrated thermal state tends to be the mixture of the two eigenvectors with the equal probabilities, which has a little tripartite quantum correlation and gives rise to the weaker thermal robustness of $T_3^{B|AC}$. In the case of frustrated thermal state ($J>0$), since the minimal negative eigenvalue $E_0$ has some intervals with other eigenvalues as shown in Fig. 5(b), the ground state $\ket{e_0}$ is dominant in a certain temperature range. Due to the ground state $\ket{e_0}$ being tripartite correlated, the robustness of $T_3^{B|AC}$ in the frustrated thermal state is stronger than that in the nonfrustrated one.

\begin{figure}
	 \epsfig{figure=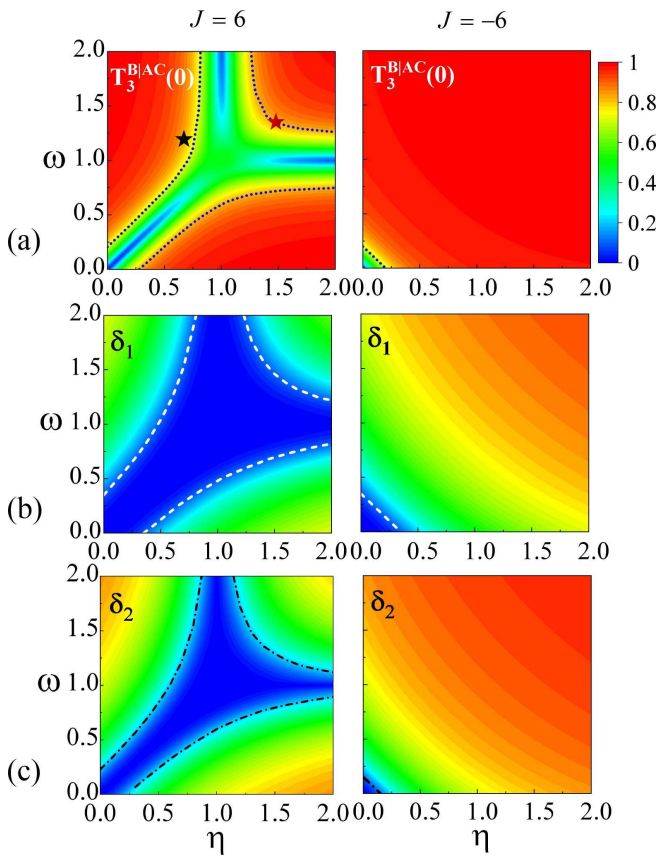,width=0.48\textwidth}
	\caption{(Color online) The MQC and its thermal effects in the spatially anisotropic system. The left column is the frustrated regime with $J=6$: (a) $T_{3}^{B|AC}$ as the function of the parameters $\omega$ and $\eta$ at zero temperature, (b) the difference $\delta_1=T_3^{B|AC}(0)-T_3^{B|AC}(0.05)$ versus the parameters $\omega$ and $\eta$, (c) the difference $\delta_2=T_3^{B|AC}(0)-T_3^{B|AC}(0.1)$ versus the parameters $\omega$ and $\eta$. Figures in the right column correspond to the nonfrustrated case with $J=-6$.}
\end{figure}

Next, we consider the spatially anisotropic triangular Ising system with the Hamiltonian $H'$ given in Eq. (8), which is completely anisotropic and has three asymmetric interactions via tuning the parameters $\omega$ and $\eta$. In order to analyze the thermal effect on the MQC in the anisotropic system, we choose the coupling strength $J=\pm 6$ away from the critical point and consider two typical values of temperature $T=0.05$ and $T=0.1$. In Fig. 6, we plot the thermal effects on $T_3^{B|AC}$ in the anisotropic system, where the left column is the frustrated case with $J=6$ and the right one corresponds to the nonfrustrated case with $J=-6$. In the left panel of Fig. 6(a), the MQC in the frustrated state is plotted as a function of the tunable parameters $\eta$ and $\omega$ at zero temperature, where the lower values (blue regions) appear at the configuration with one asymmetric interaction and higher values (red regions) appear at the configuration with the strong anisotropic interaction (the black dotted line gives the boundary for the MQC being 0.7). In the left panel of Fig. 6(b), the difference $\delta_1=T_3^{B|AC}(0)-T_3^{B|AC}(0.05)$ of the MQCs between temperatures $T=0$ and $T=0.05$ is plotted as a function of parameters $\eta$ and $\omega$, which indicates that the MQC in the systems tending to the isotropic or weak anisotropic configurations (blue region) is more thermally robust (the white dashed lines give the boundary for the difference value  $\delta_1=0.1$). Along with the increasing of temperature, the robust region for the MQC decreases as shown in the left panel of Fig. 6(c) where the dot-dashed lines indicate the boundary for the difference value $\delta_2=T_3^{B|AC}(0)-T_3^{B|AC}(0.1)=0.2$.

According to the left column of Fig. 6, we can obtain that the MQC, thermal robustness and anisotropic interactions in the frustrated Ising spins are closely related. The stronger the anisotropy of interactions, the higher the value of the MQC, but the weaker the thermal robustness of the MQC. Therefore, there exists a three-way trade-off relation among the high tripartite quantum correlation, strong thermal robustness, and the spatial anisotropy of the interactions in the frustrated regime of Ising system. In certain tasks of quantum information processing where the high MQC with strong thermal robustness is required, one can tune the interactions of frustrated spins to a proper configuration with the relatively weak anisotropy. As shown in the left panel of Fig. 6(a), the red star and black star mark two typical configurations $1:1.4:1.4$ and $1:1.18:0.72$, which have the relatively high MQC and strong thermal robustness at the same time.

In the right column of Fig. 6, the MQC for the case of nonfrustrated regime with the coupling strength $J=-6$ is plotted. The tripartite correlation $T_3^{B|AC}$ at zero temperature is plotted as a function of parameters $\eta$ and $\omega$ as shown in the right panel of Fig. 6(a), where the anisotropic parameters have little influence on the MQC except for the region with $\omega$ and $\eta$ much less than 1. The reason is that the ferromagnetic interactions make the ground state tend to the GHZ state, and the quantum state tends to a bipartite separable state in the left bottom quarter with very small values of $\omega$ and $\eta$. In the right ones of Figs.~6(b) and~6(c), the differences $\delta_1$ and $\delta_2$ for thermal robustness are plotted along with the two parameters, where we can see that the strength of two anisotropic parameters can dramatically enhance the decay rate of the MQC, although they hardly affect the initial MQC at zero temperature. In general, the tunable anisotropy in the nonfrustrated regime cannot provide the high MQC and strong thermal robustness simultaneously.

\section{discussion and conclusion}

Along with the experimental progress on manipulating the coupling among multipartite quantum systems, the tunable spatially anisotropic triangular structure can be realized in various physical platforms, such as trapped ions \cite{kim10nat}, cold atoms \cite{Struck11sci}, artificial-spin-ice heterosystem \cite{wang18natnano}, superconducting systems \cite{niskanen07sci}, and so on. Since the MQC is an important resource for quantum computation and quantum simulation, it is desirable to realize the multipartite correlation modulation in a practical system.

According to the previous analysis in Sec. II, we know that the tuning of spatial anisotropy of interactions provides us an effective strategy for the control of tripartite quantum correlations in the ground state of the frustrated regime, and the configuration with one asymmetric interaction has the ability to modulate multipartite correlation in a large range as shown in Fig.~1(c). Here, we discuss an experimental scheme for the MQC modulation by the spatially anisotropic coupling in the system of cold atoms. In Ref. \cite{Struck11sci}, the authors utilized the motional degrees of freedom of atoms trapped in a triangular optical lattice to simulate frustrated magnetism. The key technology in their experiment is the independent tuning of the nearest-neighbor coupling $J$ and $J'$ by introducing a fast oscillation of the lattice \cite{eckardt10epl}. The atoms at each site $i$ of the lattice have a local phase $\theta_i$, which can be regarded as a classical vector spin $\textbf{S}_i=[\mbox{cos}(\theta_i), \mbox{sin}(\theta_i)]$. The ground state of this system corresponds to the minimum of the energy \cite{Struck11sci}
\begin{equation}\label{13}
    E(\{\theta_i\})=-\sum_{\langle i, j\rangle} J_{i j} \mathbf{S}_{i} \cdot \mathbf{S}_{j},
\end{equation}
where the sum extends over all pairs with the nearest-neighbor coupling, and the coefficient $J_{ij}$ has the meaning of coupling strength between the $i$th and $j$th sites. In the triangular configuration simulating Ising model, one can choose $J_{AB}=J_{BC}=J$ and $J_{BC}=J'$. When the couplings $J_{ij}$ are negative, the system is frustrated. In the experimental system, the sign and magnitude of $J_{ij}$ can be tuned independently via an elliptical shaking of the lattice, which provides a tunable anisotropic parameter $\eta=J/J'$. In Ref. \cite{Struck11sci}, the experimental data can cover the region $(0.16, 1.8)$ for the spatially anisotropic parameter $\eta$, which can realize an effective modulation for the MQC from 0.1 to 0.97.

In comparison to the Ising system with one asymmetric parameter, the tunable triangular Ising model with completely spatially anisotropic interactions has more configurations with high MQC in the ground state of frustrated regime as shown in Fig. 4. In experiment, the triangular structure with three asymmetric couplings can be realized via an L-shaped three-qubit superconductor platform \cite{Grosz11prb}. Similar Ising couplings can also be simulated by a triple quantum dot system \cite{Seo13prl} and a cavity-QED system with the sign-changing atomic interactions \cite{Guo19prl}. Very recently, a quantum network with two nodes was realized in a superconducting system \cite{zhong21nat}, where each node is composed of three coupled qubits, and this experiment confirms the deterministic transfer of tripartite GHZ state from one node to the other node. In our work, the tunable Ising system with the triangular configuration can provide many kinds of multipartite correlated ground states and the MQC can be modulated effectively by the anisotropic couplings, which make this triangular structure become a potential building block for the quantum information processing in more complex quantum networks on the superconducting platform.

In Sec. III, we analyze the MQC in the thermal state of the tunable anisotropic Ising model. As shown in Fig. 6(b), although the MQC in the nonfrustrated ground state is high at zero temperature, it is fragile at finite temperatures and the anisotropic coupling accelerates the thermal decay of the MQC. However, in the frustrated regime, we find that there exists a three-way trade-off relation among high MQC, strong thermal robustness and the spatially anisotropic couplings. In Fig. 6(a), the red and black stars mark two typical anisotropic configurations, which can possess the relatively high MQC and well thermal robustness at the same time. In Fig.7, for the two marked configurations, we plot the change of tripartite quantum correlation $T_3^{B|AC}$ along with temperature $T$ where the coupling strength is chosen to be $J=\pm 6$. As shown in the figure, the spatial anisotropy obviously enhances the initial MQC to about 0.8 (the dot-dashed lines) in both the two configurations when $J$ is positive, in comparison with the value 0.5 for the isotropic case (the black dotted line in Fig. 5(a)). Meanwhile, we can see that the MQCs in the frustrated regime have more stronger thermal robustness than that in the nonfrustrated one (the solid lines).

\begin{figure}
	 \epsfig{figure=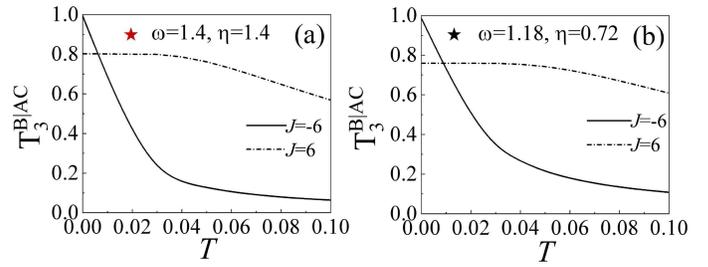,width=0.5\textwidth}
	\caption{(Color online) The tripartite quantum correlation $T_{3}^{B|AC}$ versus temperature for two typical anisotropic configurations: (a) $\omega=1.4$ and $\eta=1.4$, (b) $\omega=1.18$ and $\eta=0.72$. The dotted curves correspond to the results of frustrated regime with $J=6$, and the solid ones the nonfrustrated regime with $J=-6$.}
\end{figure}

In conclusion, we have studied the MQC in a tunable Ising system with a triangular configuration and presented a set of effective tools ($T_3$ and our newly defined susceptibility $\chi^{T_3}$) to characterize the frustrated and nonfrustrated regimes of the ground state. Meanwhile, it is revealed that the MQC in the ground state of frustrated regime can be effectively modulated by the spatially anisotropic couplings, which is feasible under current experimental technologies. Furthermore, we analyzed the MQC of thermal states in the tunable anisotropic Ising system, where we found that there exists a three-way trade-off relation among the MQC, the thermal robustness and the anisotropic coupling. In particular, one can obtain the relatively high MQC and strong thermal robustness at the same time via tuning the spatially anisotropic interactions in the frustrated Ising spins. Experimentally, the tunable triangular lattice can be utilized as a basic building unit for constructing complex multipartite quantum networks \cite{wehner18sci} and served as a potential platform for detecting quantum nonlocality in many-body systems \cite{luo18prl,renou19prl,tejada21prl}.

\begin{acknowledgments}
This work was supported by the NSF-China (Grants Nos. 11575051, 11904078 and 12105074), Hebei NSF (Grants Nos. A2021205020, A2019205263 and A2019205266), and project of China Postdoctoral Science Foundation (Grant No. 2020M670683). JR and L-HR were also funded by Science and Technology Project of Hebei Education Department (Grant No. QN2019092) and Hebei Normal University (Grants No. L2019B07).
\end{acknowledgments}

\appendix
\section{the anisotropic triangular Ising model with one tunable parameter}

In Sec.~II of the main text, we study the anisotropic triangular Ising model with one tunable parameter $\eta$, for which the ground state $\ket{\psi_0}_{ABC}$ in Eq.~(2) has the form
\begin{eqnarray}
	\ket{\psi_0}=(\alpha |001\rangle+
	\gamma |010\rangle+\alpha |100\rangle+\chi |111\rangle)/K_{0},
\end{eqnarray}
where the coefficients are
\begin{eqnarray}
	\alpha&=&J(2 J+2 J E_{0}+(2 E_{0}+E_{0}^{2}-3)\eta \nonumber\\ &&-J^{2}\eta\left(\eta^{2}-2\right)),\nonumber\\
	\gamma&=&J\left(E_{0}+J \eta-1\right)\left(E_{0}+J \eta+3\right),\nonumber\\
	\chi&=&J\left(E_{0}+J \eta-1\right)^{2},\nonumber
\end{eqnarray}
with the corresponding ground state energy $E_0$ is
\begin{eqnarray}
	E_{0}&=&-\sqrt{\frac{p_{1}}{3}} (\cos \theta_1+\sqrt{3}\sin \theta_1)- \frac{a_1}{3},
\end{eqnarray}
in which the parameters $p_1$, $\theta_1$ and $a_1$ are the function of $J$ and $\eta$, and their expressions will be given in Eq. (A7) of this Appendix.

For the tunable system with one asymmetric parameter, its thermal state can be expressed as the form of the mixture of eigenstates as shown in Eq.~(10). Beside the ground state $\ket{\psi_0}$, we can further solve its other eigenstates and corresponding energies. After some derivation, we can obtain
\begin{eqnarray}
	\ket{\psi_1}&=&(\alpha_{1} |000\rangle+ |011\rangle+\delta_{1} |101\rangle+ |110\rangle)/K_{1},\nonumber\\
	 \ket{\psi_2}&=&(|110\rangle-|011\rangle)/\sqrt{2},\nonumber\\
	 \ket{\psi_3}&=&(|100\rangle-|001\rangle)/\sqrt{2},\\
	\ket{\psi_4}&=&(\alpha_{4} |000\rangle+ |011\rangle+\delta_{4} |101\rangle+ |110\rangle)/K_{4},\nonumber\\
	\ket{\psi_5}&=&(\alpha_{5} |000\rangle+ |011\rangle+\delta_{5} |101\rangle+ |110\rangle)/K_{5},\nonumber\\
	\ket{\psi_6}&=&(\alpha_{6} |001\rangle+
	\gamma_{6} |010\rangle+\alpha_{6} |100\rangle+\chi_{6} |111\rangle)/K_{6}\nonumber,\\
	\ket{\psi_7}&=&(\alpha_{7} |001\rangle+
	\gamma_{7} |010\rangle+\alpha_{7} |100\rangle+\chi_{7} |111\rangle)/K_{7}\nonumber,
\end{eqnarray}
and the corresponding eigenstate energies
\begin{equation}
	\begin{split}
		&E_{1}=-\sqrt{\frac{p_{2}}{3}} (\cos \theta_2+\sqrt{3}\sin \theta_2)- \frac{a_2}{3}, \\
		&E_{2}=E_{3}=a_2,\\
		&E_{4}=2 \sqrt{\frac{p_{2}}{3}} \cos \theta_2 - \frac{a_2}{3},\\
		&E_{5}=-\sqrt{\frac{p_{2}}{3}} (\cos \theta_2-\sqrt{3}\sin \theta_2)- \frac{a_2}{3}, \\
		&E_{6}=2 \sqrt{\frac{p_{1}}{3}} \cos \theta_1 - \frac{a_1}{3}, \\
		&E_{7}=-\sqrt{\frac{p_{1}}{3}}  (\cos \theta_1-\sqrt{3}\sin \theta_1)- \frac{a_1}{3}.
	\end{split}
\end{equation}
In Eq.~(A3) the $K_i$s are normalization coefficients for the corresponding eigenstate $\ket{\psi_i}$, and when $i=1,4,5$ the amplitudes for corresponding eigenstates are
\begin{eqnarray}
	 \alpha_{i}&=&\frac{\left(E_{i}+1\right)\left(E_{i}+1-J \eta\right)-2 J^{2}}{J\left(E_{i}+1+J \eta\right)},\nonumber\\
	\delta_{i}&=&\frac{E_{i} \eta+2 J+\eta-J \eta^{2}}{E_{i}+1+J \eta},
\end{eqnarray}
and in the case $i=6,7$ we have the amplitudes
\begin{eqnarray}
	\alpha_{i}&=&J(2 J+2 J E_{j}+(2 E_{j}+E_{j}^{2}-3)\eta \nonumber\\ &&-J^{2}\eta\left(\eta^{2}-2\right)),\nonumber\\
	\gamma_{i}&=&J\left(E_{j}+J \eta-1\right)\left(E_{j}+J \eta+3\right),\nonumber\\
	\chi_{i}&=&J\left(E_{j}+J \eta-1\right)^{2}.
\end{eqnarray}
The parameters for the eigenstate energies shown in Eq.~(A4) have the forms
\begin{eqnarray}
a_1&=&1-J \eta,\nonumber\\
a_2&=&-1-J \eta,\nonumber\\
p_1 &=& -b_1+\frac{a_1^2}{3},\nonumber\\
p_2 &=& -b_2+\frac{a_2^2}{3}, \nonumber\\
\theta_1 &=& (\arccos[\frac{3q_1 \sqrt{3p_1}}{2p_1^2}])/3,\nonumber\\
\theta_2 &=& (\arccos[\frac{3q_2 \sqrt{3p_2}}{2p_2^2}])/3,
\end{eqnarray}
with
\begin{eqnarray}
	q_1 &=& -c_1-\frac{2a_1^3}{27}+\frac{a_1 b_1}{3}, \nonumber\\
    q_2 &=& -c_2-\frac{2a_2^3}{27}+\frac{a_2 b_2}{3},\nonumber\\
	b_1&=&-5-4J^2-2J \eta-J^2 \eta^2,\nonumber\\
	b_2&=&-5-4J^2+2J \eta-J^2 \eta^2,\nonumber\\
    c_1&=&3-4J^2+3J \eta-4J^3 \eta+J^2 \eta^2+J^3 \eta^3,\nonumber\\
	c_2&=&-3+4J^2+3J \eta-4J^3 \eta-J^2 \eta^2+J^3 \eta^3.	
\end{eqnarray}

In Sec.~III of main text, we studied the thermal state of an isotropic triangular Ising system, for which the eigenvectors and their corresponding energies can be obtained by the above results via taking the tunable asymmetric parameter $\eta=1$.

\section{the ground state and MQC for the Ising model with three different couplings}

In Eq.~(8) of main text, the Hamiltonian $H'$ characterizes the anisotropic Ising model with three different couplings, which is realized by two tunable parameters $\omega$ and $\eta$. As two typical examples, the tunable parameter $\omega$ is chosen to be $0.8$ and $1.2$. When the parameter $\omega=0.8$, the amplitudes shown in Eq.~(9) can be written as
\begin{eqnarray}
	\xi &=& 125\left(\mathcal{E}_{0}-1\right)^{2}\left(\mathcal{E}_{0}+3\right)200 J^{3} \eta\nonumber\\
	&&-5 J^{2}\left(41 \mathcal{E}_{0}+25 \eta^{2} \mathcal{E}_{0}-25 \eta^{2}+59\right), \nonumber\\
	\zeta &=& 250 J^{2}\left(\mathcal{E}_{0}+1\right) \eta+4 J^{3}\left(25 \eta^{2}+9\right)\nonumber\\
	&&+100 J\left(\mathcal{E}_{0}^{2}+2 \mathcal{E}_{0}-3\right), \nonumber\\
	\delta &=& 200 J^{2}\left(\mathcal{E}_{0}+1\right)+5 J^{3}\left(41 \eta-25 \eta^{3}\right)\nonumber\\
	&&+125\left(\mathcal{E}_{0}^{2}+2 \mathcal{E}_{0}-3\right) J \eta, \nonumber\\
	\tau &=& 125 J\left(\mathcal{E}_{0}-1\right)^{2}+200 J^{2}\left(\mathcal{E}_{0}-1\right)\eta\nonumber\\
	&&+5 J^{3}\left(25 \eta^{2}-9\right),
\end{eqnarray}
and the three-qubit quantum correlation of the corresponding  ground state can be derived, which has the form
\begin{eqnarray}
	T_{3}^{B|AC}=
	\begin{cases}
		\sqrt{2} \mathcal{K}_{0}^{2} \sqrt{f_{0}-\left(\zeta^{2}-\delta^{2}\right)^{2}}, & {J<\sqrt[3]{5 / 4 \eta}}\\
		\mathcal{K}_{0} \sqrt{ f_{1}-\frac{1}{4}\left(w_{0}-\frac{2}{\mathcal{K}_{0}^{2}}\right)^{2}}, & {J>\sqrt[3]{5 / 4 \eta}}
	\end{cases}
\end{eqnarray}
where we utilized the bipartite negativities
\begin{eqnarray}
	N_{B|AC}&=& 2 \mathcal{K}_{0}^{2} \sqrt{\xi^{2}+\delta^{2}} \sqrt{\zeta^{2}+\tau^{2}}, \nonumber\\
	N_{AB}&=& \mathcal{K}_{0}^{2}\left(t_{0}-\zeta^{2}-\delta^{2}\right),
\end{eqnarray}
and
\begin{eqnarray}
	N_{BC}=
	\begin{cases}
		 \mathcal{K}_{0}^{2}\left(n_{0}-\xi^{2}-\zeta^{2}\right), & {J<\sqrt[3]{5 / 4 \eta}}\\
		\frac{1}{2} \mathcal{K}_{0}^{2} w_{0}-1, & {J>\sqrt[3]{5 / 4 \eta}},
	\end{cases}
\end{eqnarray}
where the normalization coefficient $\mathcal{K}_0=\sqrt{\xi^2+\zeta^2+\delta^2+\tau^2}$.

In the expressions of amplitudes shown in Eq.~(B1), the ground state energy $\mathcal{E}_{0}$ can be written as
\begin{equation}
	\mathcal{E}_{0}=\left\{
	\begin{array}{rcl}
		 (-\sqrt{l_{1}}-\sqrt{l_{2}}-\sqrt{l_{3}}) / 20, & & {J<\sqrt[3]{5 / 4 \eta}}\\
		 (-\sqrt{l_{1}}+\sqrt{l_{2}}-\sqrt{l_{3}}) / 20, & & {J>\sqrt[3]{5 / 4 \eta}}
	\end{array} \right.
\end{equation}
which is a piece-wise function with the parameters
\begin{eqnarray}
	l_{1} &=& 2 \sqrt{\frac{u_{3}}{3}} \cos \theta_3-\frac{8 a_{3}}{3}, \nonumber\\
	l_{2} &=& -\sqrt{\frac{u_{3}}{3}} (\cos \theta_3+\sqrt{3}\sin \theta_3)- \frac{8a_3}{3},\nonumber\\
	l_{3} &=& -\sqrt{\frac{u_{3}}{3}} (\cos \theta_3-\sqrt{3}\sin \theta_3)- \frac{8a_3}{3},
\end{eqnarray}
in which
\begin{eqnarray}
	a_{3} &=& -150-82 J^{2}-50 J^{2} \eta^{2},\nonumber\\
	u_{3} &=& -16 a_{3}^{2}+64 c_{3}+\frac{64 a_{3}^{2}}{3},\nonumber\\
	\theta_3 &=& (\arccos[\frac{3v_3 \sqrt{3u_3}}{2u_3^2}])/3,
\end{eqnarray}
and
\begin{eqnarray}
	c_{3} &=&25 J^{2}\left(50 \eta^{2}-82 J^{2} \eta^{2}+25 J^{2} \eta^{4}+82\right)\nonumber\\
	&&+81 J^{4}-1875,\nonumber\\
	v_{3} &=& 64 b_{3}^{2}-\frac{1024 a_{3}^{2}}{27}-\frac{128 a_{3}^{3}+512 a_{3} c_{3}}{3}, \nonumber\\
	b_{3} &=& 1000-800 J^{3} \eta.
\end{eqnarray}

In the expressions of MQC and related bipartite negativities shown in Eqs. (B2)-(B4), the parameters $f_0$, $f_1$, $w_0$, $t_0$, and $n_0$ are

\begin{eqnarray}
	f_{0} &=&\left( \xi^{2}+ \zeta^{2}\right) n_{0}+ \left( \zeta^{2}+ \delta^{2}\right) t_{0}-\left(\xi^{2}-\zeta^{2}\right)^{2}, \nonumber\\
	f_{1} &=&4 \mathcal{K}_{0}^{2}\left(\xi^{2}+\delta^{2}\right)
\left(\zeta^{2}+\tau^{2}\right)-\mathcal{K}_{0}^{2}\left(\delta^{2}+\zeta^{2}-t_{0}\right)^{2},\nonumber\\
	t_{0} &=& \sqrt{\left(\zeta^{2}-\delta^{2}\right)^{2}+4 \xi^{2} \tau^{2}}, \nonumber\\
	n_{0} &=& \sqrt{\left(\xi^{2}-\zeta^{2}\right)^{2}+4 \delta^{2} \tau^{2}},\nonumber\\
	w_{0} &=& \xi^{2}+\zeta^{2}+\delta^{2}+\tau^{2}+n_{0}+s_{0}\nonumber\\
	 &+&\left|\xi^{2}+\zeta^{2}-n_{0}\right|+\left|\delta^{2}+\tau^{2}-s_{0}\right|,
\end{eqnarray}
in which $s_{0}=\sqrt{\left(\delta^{2}-\tau^{2}\right)^{2}+4 \xi^{2} \zeta^{2}}$.

When the tunable parameter is chosen to be $\omega=1.2$, the corresponding coefficients in Eq. (9) of the main text can be written as
\begin{eqnarray}
	\xi^{\prime} &=& 125\left(\mathcal{E}_{0}^{\prime}-1\right)^{2}\left(\mathcal{E}_{0}^{\prime}+3\right)-300 J^{3} \eta\nonumber\\&-&5 J^{2}\left(61 \mathcal{E}_{0}^{\prime}+25 \eta^{2} \mathcal{E}_{0}^{\prime}-25 \eta^{2}+39\right), \nonumber\\
	\zeta^{\prime} &=& 250 J^{2}\left(\mathcal{E}_{0}^{\prime}+1\right) \eta+2 J^{3}\left(75 \eta^{2}-33\right)\nonumber\\&+&150 J\left(\mathcal{E}_{0}^{\prime 2}+2 \mathcal{E}_{0}^{\prime}-3\right),\nonumber\\
	\delta^{\prime} &=& 300 J^{2}\left(\mathcal{E}_{0}^{\prime}+1\right)+5 J^{3}\left(61 \eta-25 \eta^{3}\right)\nonumber\\&+&125\left(\mathcal{E}_{0}^{\prime 2}+2 \mathcal{E}_{0}^{\prime}-3\right) J \eta, \nonumber\\
	\tau^{\prime} &=& 125 J\left(\mathcal{E}_{0}^{\prime}-1\right)^{2}+300 J^{2}\left(\mathcal{E}_{0}^{\prime}-1\right) \eta\nonumber\\&+&5 J^{3}\left(25 \eta^{2}+11\right).
\end{eqnarray}
According to the ground state of the anisotropic system, we can further calculate the negativities $N_{B|AC}$, $N_{AB}$ and $N_{BC}$, and then derive the MQC which has the form
\begin{eqnarray}
	T_{3}^{B|AC}=
	\begin{cases}
		\sqrt{2} \mathcal{K}_{0}^{\prime 2} \sqrt{f_{0}^{\prime}-\left(\zeta^{\prime 2}-\delta^{\prime 2}\right)^{2}}, & {J<\sqrt[3]{5 / 6 \eta}}\\
		\mathcal{K}_{0}^{\prime} \sqrt{ f_{1}^{\prime}-\frac{1}{4}\left(w^{\prime}-\frac{2}{\mathcal{K}_{0}^{\prime 2}}\right)^{2}}, & {J>\sqrt[3]{5 / 6 \eta}}
	\end{cases}
\end{eqnarray}

In the expressions of the coefficients in Eq.~(B10), the ground state energy is
\begin{equation}
	\mathcal{E}_{0}^{\prime}=\left\{
	\begin{array}{rcl}
		 (-\sqrt{l_{4}}-\sqrt{l_{5}}-\sqrt{l_{6}}) / 20, & & {J<\sqrt[3]{5 / 6 \eta}}\\
		 (-\sqrt{l_{4}}+\sqrt{l_{5}}-\sqrt{l_{6}}) / 20, & & {J>\sqrt[3]{5 / 6 \eta}}
	\end{array} \right.
\end{equation}
where the parameters are
\begin{eqnarray}
	l_{4} &=& 2 \sqrt{\frac{u_{4}}{3}}\cos \theta_4 -\frac{8 a_{4}}{3}, \nonumber\\
	l_{5} &=& -\sqrt{\frac{u_{4}}{3}} (\cos \theta_4 +\sqrt{3}\sin \theta_4 )- \frac{8a_4}{3},\nonumber\\
	l_{6} &=& -\sqrt{\frac{u_{4}}{3}} (\cos \theta_4 -\sqrt{3}\sin \theta_4 )- \frac{8a_4}{3},
\end{eqnarray}
with
\begin{eqnarray}
	a_{4} &=& -150-122 J^{2}-50 J^{2} \eta^{2}, \nonumber\\
	u_{4} &=& -16 a_{4}^{2}+64 c_{4}+\frac{64 a_{4}^{2}}{3}, \nonumber\\
	\theta_4 &=& (\arccos[\frac{3v_4 \sqrt{3u_4}}{2u_4^2}])/3,
\end{eqnarray}
and
\begin{eqnarray}
	c_{4} &=&25 J^{2}\left(50 \eta^{2}-82 J^{2} \eta^{2}+25 J^{2} \eta^{4}+82\right)\nonumber\\
	&&+121 J^{4}-1875, \nonumber\\
	v_{4} &=& 64 b_{4}^{2}-\frac{1024 a_{4}^{2}}{27}-\frac{128 a_{4}^{3}-512 a_{4} c_{4}}{3}, \nonumber\\
	b_{4} &=& 1000-1200 J^{3} \eta.
\end{eqnarray}

In the expression of the tripartite quantum correlation $T_{3}^{B|AC}$ shown in Eq. (B11), the related parameters are
\begin{eqnarray}
	f_{0}^{\prime}&=& \left( \xi^{\prime 2}+ \zeta^{\prime 2}\right) n^{\prime}+ \left( \zeta^{\prime 2}+ \delta^{\prime 2}\right) t^{\prime}-\left(\xi^{\prime 2}-\zeta^{\prime 2}\right)^{2}, \nonumber\\
	f_{1}^{\prime}&=&4 \mathcal{K}_{0}^{\prime 2}\left(\xi^{\prime 2}+\delta^{\prime 2}\right)\left(\zeta^{\prime 2}+\tau^{\prime 2}\right)\nonumber\\
	&&-\mathcal{K}_{0}^{\prime 2}\left(\xi^{\prime 2}+\zeta^{\prime 2}-n^{\prime}\right)^{2}, \nonumber\\
	w^{\prime}&=&\xi^{\prime 2}+\zeta^{\prime 2}+\delta^{\prime 2}+\tau^{\prime 2}+t^{\prime}+r^{\prime}\nonumber\\
	&&+\left|\xi^{\prime 2}+\tau^{\prime 2}-r^{\prime}\right|+\left|\zeta^{\prime 2}+\delta^{\prime 2}-t^{\prime}\right|,
\end{eqnarray}
in which
\begin{eqnarray}
	n^{\prime}&=&\sqrt{\left(\xi^{\prime 2}-\zeta^{\prime 2}\right)^{2}+4 \delta^{\prime 2} \tau^{\prime 2}},\nonumber\\
	t^{\prime}&=&\sqrt{\left(\zeta^{\prime 2}-\delta^{\prime 2}\right)^{2}+4 \xi^{\prime 2} \tau^{\prime 2}}, \nonumber\\
	r^{\prime}&=&\sqrt{\left(\xi^{\prime 2}-\tau^{\prime 2}\right)^{2}+4 \zeta^{\prime 2}\delta^{\prime 2}}.
\end{eqnarray}


\begin{thebibliography}{99}
	
	
	
\bibitem{amico08rmp} L. Amico, R. Fazio, A. Osterloh, and V. Vedral, Entanglement in many-body systems, Rev. Mod. Phys. \textbf{80}, 517 (2008).
	
\bibitem{horod09rmp} R. Horodecki, P. Horodecki, M. Horodecki, and K. Horodecki, Quantum entanglement, Rev. Mod. Phys. \textbf{81}, 865 (2009).

\bibitem{modi12rmp} K. Modi, A. Brodutch, H. Cable,T. Paterek, and V. Vedral, The classical-quantum boundary for correlations: Discord and related measures, Rev. Mod. Phys. \textbf{84}, 1655 (2012).

\bibitem{sach99book}  S. Sachdev, Quantum phase transitions (Cambridge University Press, Cambridge, UK, 1999).

\bibitem{chiara18rpp} G. De Chiara and A. Sanpera, Genuine quantum correlations in quantum many-body systems: a review of recent progress, Rep. Prog. Phys. \textbf{81}, 074002 (2018).


\bibitem{binder86rmp} K. Binder and A. P. Young, Spin glasses: Experimental facts, theoretical concepts, and open questions, Rev. Mod. Phys. \textbf{58}, 801 (1986).


\bibitem{Yoshida15np}	Y. Yoshida, H. Ito, M. Maesato, Y. Shimizu, H. Hayama, T. Hiramatsu, Y. Nakamura, H. Kishida, T. Koretsune, C. Hotta, and G. Saito, Spin-disordered quantum phases in a quasi-one-dimensional triangular lattice, Nat. Phys. \textbf{11}, 679 (2015).
	
\bibitem{Zhang16prl}	X.-F. Zhang, S. Hu, A. Pelster, and S. Eggert, Quantum Domain Walls Induce Incommensurate Supersolid Phase on the Anisotropic Triangular Lattice, Phys. Rev. Lett. \textbf{117}, 193201 (2016).

\bibitem{Ito16prb} H. Ito, T. Asai, Y. Shimizu, H. Hayama, Y. Yoshida, and G. Saito, Pressure-induced superconductivity in the antiferromagnet $\rm \kappa$-$\rm (ET)_2CF_3SO_3$ with
quasi-one-dimensional triangular spin lattice, Phys. Rev. B \textbf{94}, 020503(R) (2016).
	
\bibitem{Keles18prl}	A. Keles and E. Zhao, Absence of Long-Range Order in a Triangular Spin System with Dipolar Interactions, Phys. Rev. Lett. \textbf{120}, 187202 (2018).
	
\bibitem{Tala20prapp}	A. Talapatra, N. Singh, and A. O. Adeyeye, Magnetic Tunability of Permalloy Artificial Spin Ice Structures, Phys. Rev. Appl. \textbf{13}, 014034 (2020).

\bibitem{weitc05pra}  T. C. Wei, D. Das, S. Mukhopadyay, S. Vishveshwara, and P. M. Goadbart, Global entanglement and quantum criticality in spin chains, Phys. Rev. A \textbf{71}, 060305(R) (2005).

\bibitem{olivei06pra}  T. R. de oliveira, G. Rigolin, and M. C. de Oliveira, Global entanglement and quantum criticality in spin chains, Phys. Rev. A \textbf{73}, 010305(R) (2006).

\bibitem{olivei06prl}  T. R. de Oliveira, G. Rigolin, M. C. de Oliveira, and E. Miranda, Multipartite entanglement signature of quantum phase transitions, Phys. Rev. Lett. \textbf{97}, 170401 (2006).
	
\bibitem{facchi10njp}  P. Facchi, G. Florio, U. Marzolino, G. Parisi, and S. Pascazio, Multipartite entanglement and frustration, New J. Phys. \textbf{12}, 025015 (2010).

\bibitem{mon10pra}	A. Montakhab and A. Asadian, Multipartite entanglement and quantum phase transitions in the one-, two-, and three-dimensional transverse-field Ising model, Phys. Rev. A \textbf{82}, 062313 (2010).

\bibitem{giampa13pra}	S. M. Giampaolo and B. C. Hiesmayr, Genuine multipartite entanglement in the XY model, Phys. Rev. A \textbf{88}, 052305 (2013).
	

\bibitem{sunzy14pra}   Z.-Y. Sun, Y.-Y. Wu, J. Xu, H.-L. Huang, B.-F. Zhan, B. Wang, and C. B. Duan, Characterization of quantum phase transition in the XY model with multipartite correlations and Bell-type inequalities, Phys. Rev. A \textbf{89}, 022101 (2014).

\bibitem{hofm14prb}	M. Hofmann, A. Osterloh, and O. G\"uhne, Scaling of genuine multiparticle entanglement close to a quantum phase transition, Phys. Rev. B \textbf{89}, 134101 (2014).

\bibitem{Bayat17prl} A. Bayat, Scaling of Tripartite Entanglement at Impurity Quantum Phase Transitions, Phys. Rev. Lett. \textbf{118}, 036102 (2017).

\bibitem{Yamasaki18pra} H. Yamasaki, A. Pirker, M. Murao, W. D\"ur, and B. Kraus, Multipartite entanglement outperforming bipartite entanglement under limited quantum system sizes, Phys. Rev. A \textbf{98}, 052313 (2018).

\bibitem{haldar20prb}	S. Haldar, S. Roy, T. Chanda, A. Sen(De), and U. Sen, Multipartite entanglement at dynamical quantum phase transitions with nonuniformly spaced criticalities, Phys. Rev. B \textbf{101}, 224304 (2020).

\bibitem{bennett96pra} C. H. Bennett, H. J. Bernstein, S. Popescu, and B. Schumacher, Concentrating partial entanglement by local operations, Phys. Rev. A \textbf{53}, 2046 (1996).

\bibitem{coffman00pra}  V. Coffman, J. Kundu, and W. K. Wootters, Distributed entanglement, Phys. Rev. A \textbf{61}, 052306 (2000).

\bibitem{osborne06prl}  T. J. Osborne and F. Verstraete, General monogamy inequality for bipartite qubit entanglement, Phys. Rev. Lett. \textbf{96}, 220503 (2006).

\bibitem{christandl04jmp}   M. Christandl and A. Winter, Squashed entanglement: An additive entanglement measure, J. Math. Phys. \textbf{45}, 829 (2004).

\bibitem{fan07pra}   Y.-C. Ou and H. Fan, Monogamy inequality in terms of negativity for three-qubit states, Phys. Rev. A \textbf{75}, 062308 (2007).


\bibitem{byw07pra}  Y.-K. Bai, D. Yang, and Z. D. Wang, Multipartite quantum correlation and entanglement in four-qubit pure states, Phys. Rev. A \textbf{76}, 022336 (2007).

\bibitem{kimjs09pra}   J. S. Kim, A. Das, and B. C. Sanders, Entanglement monogamy of multipartite higher-dimensional quantum systems using convex-roof extended negativity, Phys. Rev. A \textbf{79}, 012329 (2009).

\bibitem{byw09pra}   Y.-K. Bai, M.-Y. Ye, and Z. D. Wang, Entanglement monogamy and entanglement evolution in multipartite systems, Phys. Rev. A \textbf{80}, 044301 (2009).

\bibitem{kimjs10jpa}   J. S. Kim and B. C. Sanders, Monogamy of multi-qubit entanglement using R\'{e}nyi entropy, J. Phys. A: Math. Theor. \textbf{43}, 445305 (2010).

\bibitem{cornelio10pra}  M. F. Cornelio and M. C. de oliveira, Strong superadditivity and monogamy of the R\'{e}nyi measure of entanglement, Phys. Rev. A \textbf{81}, 032332 (2010).

\bibitem{bxw14prl}  Y.-K. Bai, Y.-F. Xu, and Z. D. Wang, General monogamy relation for the entanglement of formation in multiqubit systems, Phys. Rev. Lett. \textbf{113}, 100503 (2014).

\bibitem{songbai16pra}   W. Song, Y.-K. Bai, M. Yang, M. Yang, and Z.-L. Cao, General monogamy relation of multiqubit systems in terms of squared R\'{e}nyi-$\alpha$ entanglement, Phys. Rev. A \textbf{93}, 022306 (2016).

\bibitem{ren21npj}  L.-H. Ren, M. Gao, J. Ren, Z. D. Wang, and Y.-K. Bai, Resource conversion between operational coherence and multipartite entanglement in many-body systems, New J. Phys. \textbf{23}, 043053 (2021).

\bibitem{vidal02pra}   G. Vidal and R. F. Werner, Computable measure of entanglement, Phys. Rev. A \textbf{65}, 032314 (2002).
	
\bibitem{rkrao13pra}   K. Rama Koteswara Rao, H. Katiyar, T. S. Mahesh, A. Sen(De), U. Sen, and A. Kumar, Multipartite quantum correlations reveal frustration in a quantum Ising spin system, Phys. Rev. A \textbf{88}, 022312 (2013).	


\bibitem{wang01pla}    X. Wang, Effects of anisotropy on thermal entanglement, Phys. Lett. A \textbf{281}, 101 (2001).

\bibitem{kamta02prl}   G. L. Kamta and A. F. Starace, Anisotropy and Magnetic Field Effects on the Entanglement of a Two Qubit Heisenberg XY Chain, Phys. Rev. Lett. \textbf{88}, 107901 (2002).

\bibitem{zhou03pra}    L. Zhou, H. S. Song, Y. Q. Guo, and C. Li, Enhanced thermal entanglement in an anisotropic Heisenberg XYZ chain, Phys. Rev. A \textbf{68}, 024301 (2003).

\bibitem{zhang05pra}   G. F. Zhang and S. S. Li, Thermal entanglement in a two-qubit Heisenberg XXZ spin chain under an inhomogeneous magnetic field, Phys. Rev. A \textbf{72}, 034302 (2005).


\bibitem{yunoki06prb}   S. Yunoki and S. Sorella, Two spin liquid phases in the spatially anisotropic triangular Heisenberg model, Phys. Rev. B \textbf{74}, 014408 (2006).
\bibitem{eckardt10epl}  A. Eckardt, P. Hauke, P. Soltan-Panahi, C.Becker, K. Sengstrock, and M. Lewenstein, Frustrated quantum antiferromagnetism with ultracold bosons in a triangular Lattice, Europhys. Lett. \textbf{89}, 10010 (2010).
\bibitem{hauke13prb}   P. Hauke, Quantum disorder in the spatially completely anisotropic triangular lattice, Phys. Rev. B \textbf{87}, 014415 (2013).



\bibitem{kim10nat} K. Kim, M. S. Chang, S. Korenblit, R. Islam, E. E. Edwards, J. K. Freericks, G. D. Lin, L. M. Duan, and C. Monroe, Quantum simulation of frustrated Ising spins with trapped ions, Nature \textbf{465}, 590 (2010).
	
\bibitem{Struck11sci} J. Struck, C. Olschlager, R. Le Targat, P. Soltan-Panahi, A. Eckardt, M. Lewenstein, P. Windpassinger, and K. Sengstock, Quantum simulation of frustrated classical magnetism in triangular optical lattices, Science \textbf{333}, 996 (2011).

\bibitem{wang18natnano}	Y.-L. Wang, X. Ma, J. Xu, Z.-L. Xiao, A. Snezhko, R. Divan, L. E. Ocola, J. E. Pearson, B. Janko, and W.-K. Kwok, Switchable geometric frustration in an artificial-spin-ice superconductor heterosystem, Nat. Nanotechnol. \textbf{13}, 560 (2018).

\bibitem{niskanen07sci}	A. O. Niskanen, K. Harrabi, F. Yoshihara, Y. Nakamura, S. Lloyd, and J. S. Tsai, Quantum coherent tunable coupling of superconducting qubits, Science \textbf{316}, 723 (2007).
    	
\bibitem{Ryazanov01prb} V. V. Ryazanov, V. A. Oboznov, A. V. Veretennikov, and A. Y. Rusanov, Intrinsically frustrated superconducting array of superconductor-ferromagnet-superconductor junctions, Phys. Rev. B \textbf{65}, 020501(R) (2001).

\bibitem{Grosz11prb}	P. Groszkowski, A. G. Fowler, F. Motzoi, and F. K. Wilhelm, Tunable coupling between three qubits as a building block for a superconducting quantum computer, Phys. Rev. B \textbf{84}, 144516 (2011).

\bibitem{Kosior13pra} A. Kosior and K. Sacha, Simulation of frustrated classical XY models with ultracold atoms in three-dimensional triangular optical lattices, Phys. Rev. A \textbf{87}, 023602 (2013).	

\bibitem{Seo13prl} M. Seo, H. K. Choi, S.-Y. Lee, N. Kim, Y. Chung, H.-S. Sim, V. Umansky, and D. Mahalu, Charge Frustration in a Triangular Triple Quantum Dot, Phys. Rev. Lett. \textbf{110}, 046803 (2013).

\bibitem{cory98prl}   D. G. Cory, M. D. Price, W. Maas, E. Knill, R. Laflamme, W. H. Zurek, T. F. Havel, and S. S. Somaroo, Experimental Quantum Error Correction, Phys. Rev. Lett. \textbf{81}, 2152 (1998).

\bibitem{knill98prl}  E. Knill and R. Laflamme, Power of One Bit of Quantum Computation, Phys. Rev. Lett. \textbf{81}, 5672 (1998).

\bibitem{rauss01prl}   R. Raussendorf and H. Briegel, A One-way Quantum Computer, Phys. Rev. Lett. \textbf{86}, 5188 (2001).

\bibitem{datta08prl} A. Datta, A. Shaji, and C. M. Caves, Quantum Discord and the Power of One Qubit, Phys. Rev. Lett. \textbf{100}, 200502 (2008).

\bibitem{lanyon08prl} B. P. Lanyon, M. Barbieri, M. P. Almeida,  and A. G. White, Experimental Quantum Computing Without Entanglement, Phys. Rev. Lett. \textbf{101}, 200501 (2008).

\bibitem{monz09prl} T. Monz, K. Kim, W. Hansel, M. Riebe, A. S. Villar, P. Schindler, M. Chwalla, M. Hennrich, and R. Blatt, Realization of the Quantum Toffoli Gate with Trapped Ions, Phys. Rev. Lett. \textbf{102}, 040501 (2009).
	
\bibitem{lanyon09natp} B. P. Lanyon, M. Barbieri, M. P. Almeida, T. Jennewein, T. C. Ralph, K. J. Resch, G. J. Pryde, J. L. O'Brien, A. Gilchrist, and A. G. White, Simplifying quantum logic using higher-dimensional Hilbert spaces, Nat. Phys. \textbf{5}, 134 (2009).

\bibitem{Carteret00jpa} H. A. Carteret and A. Sudbery, Local symmetry properties of pure three-qubit states, J. Phys. A. Math. Gen. \textbf{33}, 4981 (2000).

\bibitem{Brun01pla} T. A. Brun and O. Cohen, Parametrization and distillability of three-qubit entanglement, Phys. Lett. A \textbf{281}, 88 (2001).

\bibitem{Li14pra} X. Li and S. Ghose, Control power in perfect controlled teleportation via partially entangled channels, Phys. Rev. A \textbf{90}, 052305 (2014).

\bibitem{plenio05prl} M. B. Plenio, Logarithmic Negativity: A Full Entanglement Monotone That is not Convex, Phys. Rev. Lett. \textbf{95}, 090503 (2005).

\bibitem{zyc98pra} K. \.{Z}yczkowski, P. Horodecki, A. Sanpera, and M. Lewenstein, Volume of the set of separable states, Phys. Rev. A \textbf{58}, 883 (1998).

\bibitem{chen16pra} J.-J. Chen, J. Cui, Y.-R. Zhang and H. Fan, Coherence susceptibility as a probe of quantum phase transitions, Phys. Rev. A \textbf{94}, 022112 (2016).

\bibitem{lou04prb} P. Lou, W.-C. Wu, and M.-C. Chang, Quantum phase transition in spin-1/2 XX Heisenberg chain with three-spin interaction, Phys. Rev. B \textbf{70}, 064405 (2004).

\bibitem{hong08prb} T. Hong, V. O. Garlea, A. Zheludev, J. A. Fernandez-Baca, H. Manaka, S. Chang, J. B. Leao, and S. J. Poulton, Effect of pressure on the quantum spin ladder material  IPA-CuCl$_{3}$, Phys. Rev. B \textbf{78}, 224409 (2008).

\bibitem{Sidorov11prb} V. A. Sidorov, V. N. Krasnorussky, A. E. Petrova, A. N. Utyuzh, W. M. Yuhasz, T. A. Lograsso, J. D. Thompson, and S. M. Stishov, High-pressure study of the phase transition in the itinerant ferromagnet CoS$_{2}$, Phys. Rev. B \textbf{83}, 060412(R) (2011).

\bibitem{Hotta18prb} C. Hotta and K. Kenichi Asano, Magnetic susceptibility of quantum spin systems calculated by sine square deformation: One-dimensional, square lattice, and kagome lattice Heisenberg antiferromagnets, Phys. Rev. B \textbf{98}, 140405(R) (2018).

\bibitem{paris04lnp}  M. Paris and J. \v{R}eh\'{a}c\v{e}k, Quantum State Estimation, Lecture Notes in Physics Vol. 649 (Sparinger, Berlin) (2004).




\bibitem{arnesen01prl}  M. C. Arnesen, S. Bose, and V. Vedral, Natural Thermal and Magnetic Entanglement in the 1D Heisenberg Model, Phys. Rev. Lett. \textbf{87}, 017901 (2001).

\bibitem{wang01pra} X. Wang, Entanglement in the quantum Heisenberg XY model, Phys. Rev. A \textbf{64}, 012313 (2001).

\bibitem{wang02pra} X. Wang, Thermal and ground-state entanglement in Heisenberg XX qubit rings, Phys. Rev. A \textbf{66}, 034302 (2002).

\bibitem{sun05njp} Z. Sun, X. Wang and Y.-Q. Li, Entanglement in dimerized and frustrated spin-one Heisenberg chains, New. J. Phys. \textbf{7}, 83 (2005).

\bibitem{abliz06pra}  A. Abliz, H. J. Gao, X. C. Xie, Y. S. Wu, and W. M. Liu, Entanglement control in an anisotropic two-qubit Heisenberg XYZ model with external magnetic fields, Phys. Rev. A \textbf{74}, 052105 (2006).

\bibitem{ww06pra}   X. Wang and Z. D. Wang, Thermal entanglement in ferrimagnetic chains, Phys. Rev. A \textbf{73}, 064302 (2006).

\bibitem{hide09prl}  J. Hide, W. Son, and V. Vedral, Enhancing the Detection of Natural Thermal Entanglement with Disorder, Phys. Rev. Lett. \textbf{102}, 100503 (2009).

\bibitem{park17prl}  Y. Park, J. Shim, S.-S. B. Lee, H.-S. Sim, Nonlocal Entanglement of 1D Thermal States Induced by Fermion Exchange Statistics, Phys. Rev. Lett. \textbf{119}, 210501 (2017).

\bibitem{Shim08prb} J. Shim, H.-S. Sim, and S.-S. B. Lee, Numerical renormalization group method for entanglement negativity at finite temperature, Phys. Rev. B \textbf{97}, 155123 (2018).

\bibitem{bera18rpp}   A. Bera, T. Das, D. Sadhukhan, S. Singha Roy, A. Sen(De) and U. Sen, Quantum discord and its allies: a review of recent progress, Rep. Prog. Phys. \textbf{81}, 024001 (2018).



\bibitem{vedral04njp} V. Vedral, Mean-field approximations and multipartite thermal correlations, New J. Phys. \textbf{6}, 22 (2004).

\bibitem{Nakata09pra} Y. Nakata, D. Markham, and M. Murao, Thermal robustness of multipartite entanglement of the one-dimensional spin-1/2 XY model, Phys. Rev. A \textbf{79}, 042313 (2009).

\bibitem{Campbell13njp} S. Campbell, L. Mazzola, G. De Chiara, T. J. G. Apollaro, F. Plastina, T. Busch, and M. Paternostro, Global quantum correlations in finite-size spin chains, New J. Phys. \textbf{15}, 043033 (2013).


\bibitem{sun19pra} Z.-Y. Sun, M. Wang, Y.-Y. Wu, and B. Guo, Multipartite nonlocality and boundary conditions in one-dimensional spin chains, Phys. Rev. A \textbf{99}, 042323 (2019).


\bibitem{Guo19prl} Y. Guo, R. M. Kroeze, V. D. Vaidya, J. Keeling, and B. L. Lev, Sign-Changing Photon-Mediated Atom Interactions in Multimode Cavity Quantum Electrodynamics, Phys. Rev. Lett. \textbf{122}, 193601 (2019).

\bibitem{zhong21nat} Y. Zhong, H.-S. Chang, A. Bienfait, E. Dumur, M.-H. Chou, C. R. Conner, J. Grebel, R. G. Povey, H. Yan, D. I. Schuster, and A. N. Cleland, Deterministic Multi-Qubit Entanglement in a Quantum Network, Nature \textbf{590}, 571 (2021).


\bibitem{wehner18sci} S. Wehner, D. Elkouss, R. Hanson, Quantum internet: A vision for the road ahead, Science \textbf{362}, 303 (2018).


\bibitem{luo18prl}  M.-X. Luo, Computationally Efficient Nonlinear Bell Inequalities for Quantum Networks, Phys. Rev. Lett. \textbf{120}, 140402 (2018).

\bibitem{renou19prl}   M. O. Renou, E. B\"{a}umer, S. Boreiri, N. Brunner, N. Gisin, and S. Beigi, Genuine Quantum Nonlocality in the Triangle Network, Phys. Rev. Lett. \textbf{123}, 140401 (2019).

\bibitem{tejada21prl} P. Contreras-Tejada, C. Palazuelos, and J. I. de Vicente, Genuine Multipartite Nonlocality Is Intrinsic to Quantum Networks, Phys. Rev. Lett. \textbf{126}, 040501 (2021).




\end{thebibliography}
\end{document}